\documentstyle[11pt]{article}\textheight 230mm\textwidth 150mm
            \newcommand{\be}{\begin{eqnarray}}
            \newcommand{\ee}{\end{eqnarray}}
            \newcommand{\eel}[1]{\label{#1}\end{eqnarray}}
\newcommand{\e}[1]{\label{e:#1}\end{eqnarray}}
     \newcommand{\eg}{{\em e.g.\ }}
            \newcommand{\ie}{{\em i.e.\ }}
            \newcommand{\ga}{{\gamma}}
 
            \newcommand{\la}{{\lambda}}
            
            \newcommand{\del}{{\delta}}

\newcommand{\cM}{{\cal{M}}}

\newcommand{\cW}{{\cal{W}}}
           \newcommand{\ra}{{\rightarrow}}

 \newcommand{\lea}{{\leftarrow}}

            \newcommand{\pet}{{\cal P}}

\newcommand{\ca}{{\cal C}}

            \newcommand{\beq}{\begin{quote}}
            \newcommand{\eq}{\end{quote}}
            \newcommand{\Om}{\Omega}
   
            \newcommand{\al}{\alpha}
            \newcommand{\ben}{\begin{enumerate}}
            \newcommand{\een}{\end{enumerate}}
            \newcommand{\bit}{\begin{itemize}}
            \newcommand{\ei}{\end{itemize}}
    	\newcommand{\nn}{\nonumber}
            \newcommand{\r}[1]{(\ref{e:#1})}
            \newcommand{\edfl}[1]{\label{#1}\end{df}}
\newcommand{\vb}{{\cal h}}
\newcommand{\hb}{{\cal i}}

\newcommand{\ve}{{\varepsilon}}

\newcommand{\dagg}{^{\dag}}

\newcommand{\bett}{{\bf 1}}
	
\def\d{\partial}
\def\cC{{\cal C}}
\def\cX{{\cal X}}
\def\cW{{\cal W}}

 \def\cA{{\cal A}}

  \def\half{{1 \over 2}}

\begin{document}
\begin{titlepage}
\noindent
G\"{o}teborg ITP 98-13\\
\\

\vspace*{5 mm}
\vspace*{35mm}
\begin{center}{\LARGE\bf Quantum Sp(2)-antibrackets and open groups}
\end{center} \vspace*{3 mm} \begin{center} \vspace*{3 mm}

\begin{center}Igor Batalin\footnote{On leave of absence from
P.N.Lebedev Physical Institute, 117924  Moscow, Russia\\E-mail:
batalin@td.lpi.ac.ru.} and Robert
Marnelius\footnote{E-mail: tferm@fy.chalmers.se}\\ \vspace*{7 mm}
{\sl Institute of Theoretical Physics\\
Chalmers University of Technology\\ G\"{o}teborg University\\
S-412 96  G\"{o}teborg, Sweden}\end{center}
\vspace*{25 mm}
\begin{abstract}
The recently presented quantum antibrackets are generalized to quantum
Sp(2)-antibrackets. For the class of commuting operators there are true quantum
versions of the classical Sp(2)-antibrackets. For arbitrary operators we have a
generalized bracket structure involving higher Sp(2)-antibrackets. It is shown
that these quantum antibrackets may be obtained from generating operators
involving operators in arbitrary involutions. A recently presented quantum
master
equation for  operators, which was proposed to encode generalized quantum
Maurer-Cartan equations for arbitrary open groups, is generalized to the Sp(2)
formalism. In these new quantum master equations  the generalized Sp(2)-brackets
appear naturally.
\end{abstract}\end{center}\end{titlepage}

\setcounter{page}{1}
\setcounter{equation}{0}
\section{Introduction.}
In \cite{Quanti} we  introduced a quantum antibracket  defined by
\be
&&(f, g)_Q\equiv\half \left([f, [Q, g]]-[g, [Q,
f]](-1)^{(\ve_f+1)(\ve_g+1)}\right),
\e{101}
where $Q$ is an odd, hermitian, and nilpotent operator ($Q^2=0$). This
expression
satisfies all desired properties of a quantum antibracket for a class of
commuting
operators provided $Q$ is such that \r{101} belongs to the same class of
commuting
operators as $f$ and $g$. The so called quantum master equation in the BV
quantization was then shown to have the general form
\be
&&Q|\phi\hb=0.
\e{102}

Remarkably enough the bracket \r{101} satisfies a consistent algebra even for
arbitrary operators $f$ and $g$. However, for arbitrary operators \r{101}
violates
Leibniz' rule, and the would be Jacobi identities couple to higher order
brackets
defined in a definite way \cite{Quanti, OG}. The 3-antibracket was
explicitly given
in \cite{OG}. For operators in  involution it is possible to construct
generating
operators for
\r{101} and all higher antibrackets \cite{Quanti, OG}. Furthermore, as was
shown in
\cite{Quanti,OG}, there is a new type of quantum master equation which seems to
encode the generalization of the Maurer-Cartan equations for arbitrary
quantum open groups.

In the present paper we generalize most of the results of
\cite{Quanti,OG} to quantum
Sp(2)-antibrackets. (The classical Sp(2)-antibrackets
appear in the Sp(2)-extended BV
quantization
\cite{sp2,sp2s,Trip1,Trip2}.) As was already mentioned in
\cite{Quanti} an Sp(2)-generalization of \r{101} involves two odd,
hermitian, and
nilpotent operators $Q^a$,
$a=1,2$, which anticommute (see \r{202} below). The general quantum master
equation \r{102} generalizes then to
\be
&&Q^a|\phi\hb=0, \quad a=1,2,
\e{103}
which indeed are natural equations since the
Sp(2) formalism is  directly related to
the so called BRST-antiBRST quantization \cite{antiB}.
(Nilpotent anticommuting
operators and  Sp(2)-antibrackets are the basic
ingredients in the Sp(2)-formalism.)

In section 2 we define and give some properties of
the quantum Sp(2)-antibrackets. We
also give the operator form of triplectic quantization
\cite{Trip1,Trip2}. In sections
3,4 and 5 we generalize the results of  \cite{OG}
to the Sp(2)-case. In section 3 we
define Lie equations for finite gauge transformations
generated by constraints in
arbitrary involutions within an Sp(2)-extended
BFV-BRST scheme. In section 4 these
Lie equations are used to define  generating
operators of the quantum Sp(2)-antibrackets
for operators in arbitrary involutions. In
section 5 we present quantum master
equations  for the integrability conditions
of the Lie equations in section 3, which
may be viewed as generalized quantum
Maurer-Cartan equations within the
Sp(2)-formalism. Finally we end the
paper in section 6 where we give some formal
properties of the quantum Sp(2)-antibrackets
in connection with triplectic
quantization. We also give the general
automorphisms of the quantum master
equations in \cite{OG} and section 5.   In
appendix A we give the defining properties
of the conventional classical
Sp(2)-antibrackets, and in appendix B we
review the basics of the Sp(2) extended
BFV-BRST scheme. It is shown how constraints
in arbitrary involutions may be embedded
in two nilpotent charges with an Sp(2)-relation.

\section{Quantum Sp(2)-antibrackets.}

The quantum Sp(2)-antibrackets are defined by
($a, b, c, \ldots=1,2$, are $Sp(2)$-indices which are raised (and lowered)
by the
Sp(2) metric $\ve^{ab}$ ($\ve_{ab}$))\footnote{Eqs.(3)-(6), (10) and (11) were
also given in \cite{Quanti}.}
\be
&&(f, g)^a_Q\equiv\half \left([f, [Q^a, g]]-[g, [Q^a,
f]](-1)^{(\ve_f+1)(\ve_g+1)}\right),
\e{201}
where the odd operators $Q^a$ satisfy
\be
&&Q^{\{a}Q^{b\}}\equiv Q^{a}Q^{b}+Q^{b}Q^{a}\equiv [Q^{a}, Q^{b}]=0.
\e{202}
These properties imply the relations
\be
&&[Q^{\{a}, (f, g)^{b\}}_Q]=([Q^{\{a}, f], g)^{b\}}_Q+(f, [Q^{\{a},
g])^{b\}}_Q(-1)^{\ve_f+1}=[[Q^{\{a},f], [Q^{b\}}, g]].
\e{2021}
The expressions \r{201} satisfy all the properties of the corresponding
classical Sp(2)-antibrackets as defined in  the appendix A except for the Jacobi
identities 4) and Leibniz' rule 5). Instead of Leibniz' rule we have
\be
&&(fg, h)_Q^a-f(g,h)_Q^a-(f,h)_Q^a g(-1)^{\ve_g(\ve_h+1)}=
\nn\\&&=\half\left([f, h][g,
Q^a](-1)^{\ve_h(\ve_g+1)}+[f, Q^a][g, h](-1)^{\ve_g}\right),
\e{203}
and instead of the Jacobi identities we have
\be
&&(f, (g,
h)^{\{a}_Q)_Q^{b\}}(-1)^{(\ve_f+1)(\ve_h+1)}+cycle(f,g,h)=\nn\\&&
=-\half[(f,g,h)_Q^{\{a},
Q^{b\}}](-1)^{(\ve_f+1)(\ve_h+1)},
\e{204}
where
\be
&&(f,g,h)^a_Q(-1)^{(\ve_f+1)(\ve_h+1)}\equiv{1\over 3}\left([(f,g)_Q^a,
h](-1)^{\ve_h+(\ve_f+1)(\ve_h+1)}+cycle(f,g,h)\right)
\e{205}
are quantum 3-antibrackets within the Sp(2) formalism. In fact, by means of
generalized Jacobi identities for these 3-antibrackets we may derive
4-antibrackets
and so on. (Similar classical higher Sp(2)-antibrackets were considered in
\cite{hanti}.) These higher antibrackets are expressed in terms of the
lower ones
and should terminate at a certain level depending on the properties of
$Q^a$. The quantum antibrackets \r{201} with $Q^a$ operators satisfying \r{202}
determine therefore a consistent, generalized scheme involving higher
antibrackets
and a modified Leibniz' rule. In sections 5 and 6, when we consider generating
operators of these brackets and  new kind of master equations,  the above
generalized scheme will appear naturally.

From \r{203} it is clear that Leibniz' rule is satisfied if we restrict
ourselves
to the class of commuting operators $f, g, h$. This class we denote by $\cM$
in the following. In terms of operators on $\cM$ the brackets \r{201} reduce to
\be
&&(f, g)^a_Q\equiv[f, [Q^a, g]]=[[f, Q^a], g], \quad \forall f, g\in\cM.
\e{206}
If we, furthermore, require the vanishing
of the 3-antibrackets \r{205} for operators
on $\cM$, then these brackets satisfy all
properties corresponding to the defining
properties of the classical Sp(2)-antibrackets
as given in  the appendix A. This means
that \r{206} then are the true quantum Sp(2)-antibrackets
for the class of commuting
operators. The vanishing of the 3-antibrackets restrict the form of the odd
operators $Q^a$ in such a way that the quantum antibrackets \r{206} of commuting
operators also belong to the same class of commuting operators. We have
\be
&&[f, [g, [h, Q^a]]]=0, \quad \forall f,g,h\in\cM \quad \Leftrightarrow
\quad (f,
g)^a_Q\in\cM, \quad \forall f,g\in\cM.
\e{207}

In order to demonstrate the existence of $Q^a$ operators satisfying \r{202} and
\r{207} we give
 an explicit representation.  Consider  a triplectic manifold
with Darboux coordinates $x^{\al}$ and
$x^*_{\al a}$,
$\al=1,\ldots,n$, where $\ve(x^*_{\al a})=\ve_\al+1$,
$\ve_\al\equiv\ve(x^\al)$ (see
\cite{sp2,Trip1,Trip2}). Consider then these
Darboux coordinates to be coordinates on a
symplectic manifold. The canonical coordinates of this symplectic manifold
are then
$\{x^{\al}, x^*_{\al a}, p_{\al}, p^{\al a}_*\}$. After quantization we
choose all
operators which depends on
$x^{\al}$ and $x^*_{\al a}$ as the class of commuting operators $\cM$. We
may then
define the quantum antibracket by
\r{201} with
$Q^a$ given by
\be
&&Q^a=p_{\al} p^{\al a}_*(-1)^{\ve_\al},
\e{208}
which obviously satisfy \r{202} and \r{207}. In this case we have
\be
&&(f, g)^a_Q=[f, [Q^a, g]]=[[f, Q^a], g]=\nn\\
&&=(-1)^{\ve_\al}[f, p_{\al}][p_*^{\al a},
g]-(-1)^{\ve_\al}[g, p_{\al}][p^{\al a}_*,
f](-1)^{(\ve_f+1)(\ve_g+1)}.
\e{209}
Since the nonzero canonical commutation relations are,
\be
&&[x^{\al}, p_{\beta}]=i\hbar\del^\al_\beta,\quad [x^*_{\al a}, p^{\beta
b}_*]=i\hbar\del^\beta_\al\del^b_a;\nn\\&&
p_{\al}=-i\hbar\d_{\al}(-1)^{\ve_{\al}},
\quad p^{\al a}_*=i\hbar\d^{\al a}_*(-1)^{\ve_\al},
\e{210}
we find
\be
&&(i\hbar)^{-2}(f, g)^a_Q=f\stackrel{\lea}{\d}_{\al}
\d_*^{\al a}g-g\stackrel{\lea}{\d}_{\al}\d_*^{\al a} f(-1)^{(\ve_f+1)(\ve_g+1)}
\e{211}
in accordance with the correspondence between quantum and classical
antibrackets as given  in \cite{Quanti}, \ie $(i\hbar)^{-2}(f,
g)^a_Q\leftrightarrow
(f, g)^a$. The wave function representation $\Delta^a$ of the Sp(2)-charges
\r{208}
follows from the equalities
\be
&&\vb x, x^*|Q^a|\phi\hb=-(i\hbar)^2\Delta^a\phi(x, x^*),
\quad \phi(x, x^*)\equiv\vb x,
x^*|\phi\hb.
\e{212}
Obviously, $\Delta^a$ are the odd differential operators
\be
&&\Delta^a=(-1)^{\ve_\al}{\d\over\d x^{\al}}{\d\over\d x^*_{\al a}},
\e{213}
which play a crucial role in the classical Sp(2) formalism \cite{sp2}.

More general brackets are obtained if we consider  general coordinates
$X^A=(x^{\al};x_{\al a}^*)$,
$\ve(X^A)\equiv\ve_A$. The most general form of the operators $Q^a$ which yield
Sp(2)-antibrackets satisfying all required properties are of the form
\be
&&Q^a=-\half\rho^{-1/2}P_A\rho E^{ABa}P_B\rho^{-1/2}(-1)^{\ve_B},
\e{214}
where $E^{ABa}(X)=-E^{BAa}(X)(-1)^{(\ve_A+1)(\ve_B+1)}$ and $\rho(X)$ are the
triplectic metric and the volume form density respectively.
$P_A$ satisfies
\be
&&[X^A, P_B]=i\hbar\del_B^A; \quad
P_A=-i\hbar\rho^{-1/2}\d_A\circ\rho^{1/2}(-1)^{\ve_A}, \quad \d_A\equiv
\d/\d X^A.
\e{215}
By means of \r{206} the quantum antibrackets \r{211} generalize here to (cf
\cite{Trip1,Trip2})
\be
&&(i\hbar)^{-2}(f, g)^a_Q=f\stackrel{\lea}{\d}_A E^{ABa}\d_{B}\, g,
\e{216}
where the commuting operators $f$ and $g$ are arbitrary functions of $X^A$.
These are the general forms of the Sp(2)-brackets.
The property \r{202} of $Q^a$ requires the tensor $E^{ABa}$ to satisfy the
cyclic
relations
\cite{Trip1,Trip2}
\be
&&E^{AD\{a}\d_DE^{BCb\}}(-1)^{(\ve_A+1)(\ve_C+1)}+cycle(A,B,C)=0,
\e{217}
which make the antibrackets \r{216}  satisfy the Jacobi identities. If momenta
$P_A$ are allowed to enter the operators
$Q^a$ more than quadratically then a nonzero contribution appears on the
right-hand
side of \r{217}. This means that we then have  nonzero 3-antibrackets \r{205}.

In triplectic quantization the general master equations are (see \cite{Trip2})
\be
&&Q^a_+|\phi\hb=0,
\e{218}
where
\be
&&Q^a_\pm\equiv-\half\rho^{-1/2}(P_A\mp F_A)\rho
E^{ABa}(P_B\mp F_B)\rho^{-1/2}(-1)^{\ve_B}.
\e{219}
$F_A$ are functions of the operators $X^A$. (This $F_A$ differ from the
$F_A$ in [6]
by a sign factor $(-1)^{(\ve_A+1)}$.) $F_A$, $E^{ABa}$,
 and $\rho$   are required to be
such that
$Q^{a}_\pm$ are hermitian and
\be
&&[Q^{a}_\pm, Q^{b}_\pm]=0, \quad F_AE^{ABa}F_B(-1)^{\ve_B}=0.
\e{220}
Note that the antibrackets defined in terms of $Q^a_\pm$ are the same as those
defined in terms of $Q^a$ in \r{214} for operators which are functions of $X^A$.
The partition function $Z$, \ie the path
integral of the gauge fixed action, is here given by
$Z=\vb \cX|\cW\hb$, where $|\cW\hb$ is the master state and $|\cX\hb$ a
gauge fixing
state both satisfying the quantum master equations \r{218}.
$|\cW\hb$  and $|\cX\hb$ have the general form
\be
&&|\cW\hb=\exp{\left\{{i\over\hbar}\cW(X)\right\}}\rho^{1/2}|0\hb_{P\pi}, \quad
|\cX\hb=\exp{\left\{-{i\over\hbar}\cX\dagg(X,\la)\right\}}\rho^{1/2}|0\hb_{P
\pi},
\e{221}
 where $\la^\al$ are
Lagrange multipliers and
$\pi_\al$ their conjugate momenta. The vacuum state $|0\hb_{P\pi}$ satisfies
$P_A|0\hb_{P\pi}=\pi_\al|0\hb_{P\pi}=0$. By means of coordinate
states
$|X,\la\hb$ satisfying the completeness relations
\be
&&\int |X,\la\hb\rho(X)dX d\la\vb X,\la|=\bett,
\e{222}
and $\vb X,\la|\rho^{1/2}| 0\hb_{P,\la}=1$,
the partition function $Z=\vb \cX|\cW\hb$ becomes
explicitly
\be
&&Z=\vb \cX|\cW\hb=\int\rho(X)dX d\la
\exp{\left\{{i\over\hbar}\biggl[\cW(X)+\cX(X,\la)\biggr]\right\}}
\e{223}
where  $\cW$ and $\cX$ in the path integral denotes
the master action and gauge fixing actions  $\vb
X,\la|\cW(X)\rho^{1/2}| 0\hb_{P,\la}$
 and $\vb X,\la|\cX(X,\la)\rho^{1/2}| 0\hb_{P,\la}$
respectively, which by
\r{218} and \r{221} satisfy the quantum
master equations given in \cite{Trip1,Trip2}.
In fact, \r{223} agrees with the general
partition function in \cite{Trip1,Trip2} and
for its  precise meaning we refer to these papers. (When \r{221} satisfies
\r{102}
with $Q$ given by (see (48) in \cite{Quanti})
\be
&&Q\equiv-\half\rho^{-1/2}P_A\rho
E^{AB}P_B\rho^{-1/2}(-1)^{\ve_B},
\e{224} then \r{223} represents the partition function for generalized BV
quantization. This generalizes appendix A in \cite{Quanti}.)

\section{Integrating open groups within  the Sp(2)-formalism.}

Finite gauge transformations are obtained by integrations of Lie equations like
\be
&&A(\phi)\stackrel{\lea}{\nabla}_\al\equiv
A(\phi)\stackrel{\lea}{\d_\al}-(i\hbar)^{-1}[A(\phi), Y_\al(\phi)]=0,
\e{401}
where $\d_\al$ is a derivative with respect to the parameter $\phi^\al$,
$\ve(\phi^\al)=\ve_\al$, and where the $\phi$-dependent operators $Y_\al$
($\ve(Y_{\al})=\ve_\al$) should be connected to the gauge generators.
Usually constraint operators in involutions like $\theta_\al$ in \r{306} in
appendix B generate general gauge transformations. However, if we consider a
formulation which is embedded in the conventional BFV-BRST formulation,
 appropriate hermitian gauge generators are of the form
\be
&&\tilde{\theta}_\al\equiv [\Om, \pet_\al]=\theta_\al+\{\mbox{\small
possible ghost dependent terms}\},
\e{402}
where $\Om$ is the BFV-BRST charge. In \cite{OG} the above reasoning  led us to
consider the operator
 $Y_\al$ in \r{401} to be of  the general form
\be
&&Y_a(\phi)=(i\hbar)^{-1}[\Om, \Om_\al(\phi)],\quad\ve(\Om_\al)=\ve_\al+1.
\e{403}
This form of $Y_\al$ implies that  a BRST invariant operator remains BRST
invariant
when
 transformed according to
\r{401}. From treatments of quasigroups we found that
 the operator $Y_\al$ should have the explicit form
\be
&&Y_\al(\phi)=\la^\beta_\al(\phi)
\theta_\beta(-1)^{\ve_\al+\ve_\beta}+\{\mbox{\small
possible ghost dependent terms}\},\quad \la^\beta_\al(0)=\del^\beta_\al,
\e{404}
where $\la^\beta_\al(\phi)$ ($\ve(\la^\beta_\al)=\ve_\al+\ve_\beta$) in
general are
operators. For quasigroups  we  could also
choose $Y_\al(0)=\tilde{\theta}_\al$. The explicit form  \r{404} implied that
$\Om_\al$ in  \r{403} should explicitly look like
\be
&&\Om_\al(\phi)=\la^\beta_\al(\phi)\pet_\beta+\{\mbox{\small possible ghost
dependent terms}\}.
\e{405}
In \cite{OG} a simple quantum  master equation  was then set up
for the operators
$\Om_\al$, which could be viewed as generalized quantum Maurer-Cartan
equations. (See
\r{b1} in section 7.)

We shall now generalize the results of \cite{OG} to the Sp(2) scheme. First we
notice that the generalization of \r{402} is given by
\be
&&\tilde{\theta}^b_{\al a}\equiv [\Om^b, \pet_{\al
a}]=\theta_\al\del^b_a+\{\mbox{\small possible ghost dependent terms}\},
\e{406}
where $\Om^a$ are the Sp(2)-charges in appendix B.
This suggests that the generalization of \r{401} is
\be
&&A(\phi)\stackrel{\lea}{\nabla}^b_{\al a}\equiv
A(\phi)\stackrel{\lea}{\d_\al}\del_a^b-(i\hbar)^{-1}[A(\phi), Y^b_{\al
a}(\phi)]=0,
\e{407}
where the operators $Y^b_{\al a}(\phi)$ ($\ve(Y^b_{\al a})=\ve_\al$) are of
the form
\be
&&Y^b_{\al
a}(\phi)=\la^\beta_\al(\phi)\theta_\beta
\del_a^b(-1)^{\ve_\al+\ve_\beta}+\{\mbox{\small
possible ghost dependent terms}\},
\e{408}
where $\la^\beta_\al(\phi)$ are the same  operators as in \r{404} satisfying
$\la^\beta_\al(0)=\del^\beta_\al$. At least for quasigroups it is natural to
expect $Y^b_{\al a}(0)=\tilde{\theta}^b_{\al a}$.
The equations \r{407} are equivalent to \r{401} with $Y_\al\equiv\half
Y^a_{\al a}$
{\em and}
\be
&&[A(\phi), T^{ab}_\al(\phi)]=0, \quad
T^{ab}_\al(\phi)\equiv\ve^{\{ac}Y^{b\}}_{\al c}(\phi).
\e{410}
Thus, in distinction to the case in the conventional BFV-BRST formulation
we have here
also to impose the boundary conditions
\be
&&[A(0), T^{ab}_\al(0)]=0.
\e{411}
These conditions restrict the class of operators $A(0)$ that may be integrated
in the Sp(2) scheme.  From  \r{408} we have
 $T^{ab}_\al(0)\equiv\ve^{\{ac}Y^{b\}}_{\al c}(0)=\{\mbox{\small
possible ghost dependent terms}\}$. If we also assume that $Y^{b}_{\al
a}(0)=\tilde{\theta}^b_{\al a}$  these ghost dependent terms may be calculated
from
$\Om^a$ in appendix B.  It then looks like
$A(0)$ is more and more restricted the more complicated and the higher the
rank of
the group.
 It is unrestricted for abelian theories, ghost independent for Lie group
theories and so on.

The integrability of the equations \r{407} require the following conditions
to be
satisfied
\be
&&0=\del_a^c\del_b^dA\left(\stackrel{\lea}{\d_\al}\stackrel{\lea}{\d_\beta}-
\stackrel{\lea}{\d_\beta}\stackrel{\lea}{\d_\al}
(-1)^{\ve_\al\ve_\beta}\right)=\nn\\&&=
(i\hbar)^{-1}[A, Y_{\al a}^c\stackrel{\lea}{\d_\beta}\del_b^d-
Y_{\beta
b}^d\stackrel{\lea}{\d_\al}\del_a^c(-1)^{\ve_\al\ve_\beta}-(i\hbar)^{-1}[Y_{
\al a}^c,
Y_{\beta b}^d]\: ].
\e{413}
Now since $A$ is restricted in general we cannot conclude that the
operators in the
second entry of the commutator are equal to zero. The only integrability
conditions independent of $A$ are
\be
&&Y_{\al \{a}^{\{c}\stackrel{\lea}{\d_\beta}\del_{b\}}^{d\}}-
Y_{\beta
\{b}^{\{d}\stackrel{\lea}{\d_\al}\del_{a\}}^{c\}}
(-1)^{\ve_\al\ve_\beta}-(i\hbar)^{-1}[Y_{\al
\{a}^{\{c}, Y_{\beta b\}}^{d\}}]=0.
\e{414}
The remaining conditions in \r{413} are then conditions on $A$. In fact,
they are
consistency conditions to \r{410} which ensure the properties
($Y_\beta\equiv\half
Y^a_{\beta a}$)
\be
&& [A(\phi), [T^{ab}_\al(\phi), T^{cd}_\beta(\phi)] \:]=0,
\nn\\&&[A(\phi), T^{ab}_\al(\phi)]\stackrel{\lea}{\d_\beta}=[A(\phi),
T^{ab}_\al(\phi)\stackrel{\lea}{\d_\beta}-(i\hbar)^{-1}[T_\al^{ab},
Y_\beta] \:]= 0.
\e{415}
Notice also that as a consequence of \r{414},
$Y_\al\equiv\half Y^a_{\al a}$ satisfies the equation
\be
&&Y_{\al}\stackrel{\lea}{\d_\beta}-
Y_{\beta
}\stackrel{\lea}{\d_\al}
(-1)^{\ve_\al\ve_\beta}-(i\hbar)^{-1}[Y_{\al
}, Y_{\beta}]= -{1\over 24}(i\hbar)^{-1}\ve_{ac}[T_\al^{ab},
T_\beta^{cd}]\ve_{bd},
\e{4151}
which together with the first equation in \r{415} guarantees the
integrability of \r{401}.

In analogy to \r{403} we propose   $Y^b_{\al a}$ to be of the general form
\be
&&Y^b_{\al a}(\phi)=(i\hbar)^{-1}[\Om^b, \Om_{\al
a}(\phi)],\quad\ve(\Om_{\al a})=\ve_\al+1,
\e{416}
which makes $Y^b_{\al a}$ Sp(2)-invariant in the sense
\be
&&[\Om^{\{a}, Y^{b\}}_{\al c}(\phi)]=0.
\e{417}
Furthermore, we have then
\be
&&[\Om^{\{c}, A] \stackrel{\lea}{\d_\al}\del_a^{b\}}=(i\hbar)^{-1}[[\Om^{\{c}, A],
Y_{\al a}^{b\}}],
\e{4171}
which implies that $[\Om^a, A(\phi)]=0$ if $[\Om^a, A(0)]=0$.  Due to the
explicit
form \r{408} of
$Y^b_{\al a}$,
$\Om_{\al a}$ in \r{416} should  look like
\be
&&\Om_{\al a}(\phi)=\la^\beta_\al(\phi)\pet_{\beta a}+\{\mbox{\small
possible ghost
dependent terms}\}.
\e{418}
 If we insert the general form
\r{416} of $Y_{\al a}^b$ into the integrability conditions \r{414},  we
find  the
following equivalent equations for
$\Om_{\al a}$
\be
&&\Om_{\al
\{a}\stackrel{\lea}{\d_\beta}\del_{b\}}^c-\Om_{\beta
\{b}\stackrel{\lea}{\d_\al}\del_{a\}}^c(-1)^{\ve_\al\ve_\beta}=\nn\\&&=
 (i\hbar)^{-2}(\Om_{\al
\{a},
\Om_{\beta b\}})^c_{\Om}-\half(i\hbar)^{-1}[\Om_{\al \beta a b}, \Om^c],
\e{419}
where $(\Om_{\al a},
\Om_{\beta b})^c_{\Om}$ are the Sp(2)-antibrackets \r{201} with
 $Q^a$ replaced by $\Om^a$. $\Om_{\al \beta a b}$ is symmetric in $a$, $b$, and
antisymmetric in $\al$, $\beta$. Due to the explicit form
 \r{418} of $\Om_{\al a}$, these equations are generalized Maurer-Cartan
equations for
$\la^\beta_\al(\phi)$ which should be equivalent to those obtained from
$\Om_\al$ in
\r{405} as given in \cite{OG}. Now the equations
\r{419} are only integrable if
$\Om_{\al a}$ and
$\Om_{\beta b}$ commute and if simultaneously the 3-antibrackets \r{205} for
the operators $\Om_{\al a}$ are zero or equivalently if
\r{207} with
$f, g ,h$ and
$Q^a$ replaced by the $\Om_{\al a}$,  $\Om_{\beta b}$, $\Om_{\ga c}$, and
$\Om^d$ are
satisfied. In this case
$\Om_{\al \beta a b}$ in \r{419} is zero. If these conditions are not
satisfied the
integrability conditions of
\r{419} lead to equivalent first order equations of $\Om_{\al \beta a b}$
and so on.
$Y_{\al a}$ is then replaced by a whole set of operators, and  the integrability
conditions \r{414} for
$Y_{\al a}^b$ are replaced by a whole set of integrability conditions.
In section 6 we
propose new simple quantum master equations for the operators  $\Om_{\al a}$,
$\Om_{\al \beta a b}$ etc.

\section{Generating operators for Sp(2)-antibrackets of operators in arbitrary
involutions.}

In \cite{Quanti,OG} it was shown that quantum antibrackets for operators in
arbitrary
involutions may be derived from a generating, nilpotent
operator $Q(\phi)$ satisfying the Lie equations \r{401}. Here we generalize this
construction to the Sp(2)-formalism. Let
$Q^a$ be nilpotent Sp(2)-charges satisfying \r{202}. We define then
$Q^a(\phi)$ with the boundary conditions $Q^a(0)=Q^a$ and \r{411} to be
solutions to
the equations
\r{407},
\ie
\be
&&Q^c(\phi)\stackrel{\lea}{\nabla}^b_{\al a}\equiv
Q^c(\phi)\stackrel{\lea}{\d_\al}\del_a^b-(i\hbar)^{-1}[Q^c(\phi), Y^b_{\al
a}(\phi)]=0.
\e{501}
This  implies
\be
&&[Q^a(\phi), Q^b(\phi)]\stackrel{\lea}{\d_\al}\del_c^d=(i\hbar)^{-1}[
[Q^a(\phi),
Q^b(\phi)], Y^d_{\al c}(\phi)],
\e{502}
which by means of the boundary conditions $Q^a(0)=Q^a$ ensures that
\be
&&[Q^a(\phi), Q^b(\phi)]=0.
\e{503}
This shows that $Q^a\,\ra\, Q^a(\phi)$ is a unitary transformation.

Following refs.\cite{Quanti, OG} we define generalized quantum antibrackets
in terms
of
$Q^c(\phi)$ according to the formula
\be
&&{(Y^{b_1}_{\al_1a_1}(\phi),
Y^{b_2}_{\al_2a_2}(\phi),\ldots,
Y^{b_n}_{\al_na_n}(\phi))'}^c_{Q(\phi)}\equiv\nn\\
&&\equiv-Q^c(\phi)
\stackrel{\lea}{\d}_{\al_1}\del_{a_1}^{b_1}
\stackrel{\lea}{\d}_{\al_2}\del^{b_2}_{a_2}\cdots
\stackrel{\lea}{\d}_{\al_n}\del^{b_n}_{a_n}(i\hbar)^{n}(-1)^{E_n},\quad
E_n\equiv
\sum_{k=0}^{\left[{n-1\over 2}\right]}\ve_{\al_{2k+1}}.
\e{5031}
Explicitly we have then \eg
\be
&&{(Y_{\al a}^c(\phi),
Y_{\beta b}^d(\phi))'}^f_{Q(\phi)}
\equiv-Q^f(\phi)\stackrel{\lea}{\d_\al}\del_a^c
\stackrel{\lea}{\d_\beta}\del_b^d(-1)^{\ve_\al}(ih)^2=\nn\\
&&=\half\left([Y_{\al a}^c(\phi),
[Q^f(\phi), Y_{\beta b}^d(\phi)]]-[Y_{\beta b}^d(\phi), [ Q^f(\phi),
Y_{\al a}^c(\phi)]](-1)^{(\ve_\al+1)(\ve_\beta+1)}\right)-\nn\\
&&-\half i\hbar[Q^f(\phi),
Y_{\al
a}^c(\phi)\stackrel{\lea}{\d_\beta}\del^d_b+Y_{\beta
b}^d(\phi)\stackrel{\lea}{\d_\al}\del^c_a(-1)^{\ve_\al\ve_\beta}](-1)^{\ve_\al}.
\e{50311}
Since $Y^b_{\al a}(\phi)$ may be split as follows
\be
&&Y^b_{\al a}(\phi)\;\Leftrightarrow\;
Y_\al(\phi)\equiv\half Y^a_{\al a}(\phi),\quad
T^{ab}_\al(\phi)\equiv\ve^{\{ac}Y^{b\}}_{\al c}(\phi),
\e{5032}
it is easily seen that \r{5031} implies that the required antibrackets are
given by
\be
&&{(Y_{\al_1}(\phi),
Y_{\al_2}(\phi),\ldots,Y_{\al_n}(\phi))'}^c_{Q(\phi)}\equiv-Q^c(\phi)
\stackrel{\lea}{\d}_{\al_1}\stackrel{\lea}{\d}_{\al_2}\cdots
\stackrel{\lea}{\d}_{\al_n}(i\hbar)^{n}(-1)^{E_n},
\e{504}
and that these antibrackets are zero if any entry is $T^{ab}_\al(\phi)$, \ie
\be
&&{(\cdots, T^{ab}_\al(\phi), \cdots)'}^a=0.
\e{5041}
Note that only $Y_\al$ in \r{5032} involves  the constraint operators
$\theta_\al$.
By means of the equations
\be
&&Q^a(\phi)\stackrel{\lea}{\d_\al}=
(i\hbar)^{-1}[Q^a(\phi), Y_{\al}(\phi)],
\e{505}
which follow from \r{501}, we have then from \r{504} in particular
\be
&&{(Y_{\al}(\phi),
Y_{\beta }(\phi))'}^a_{Q(\phi)}\equiv-Q^a(\phi)\stackrel{\lea}{\d_\al}
\stackrel{\lea}{\d_\beta}(-1)^{\ve_\al}(ih)^2=\nn\\
&&=\half\left([Y_{\al}(\phi),
[Q^a(\phi), Y_{\beta}(\phi)]]-[Y_{\beta}(\phi), [ Q^a(\phi),
Y_{\al}(\phi)]](-1)^{(\ve_\al+1)(\ve_\beta+1)}\right)-\nn\\
&&-\half i\hbar[Q^a(\phi),
Y_{\al}(\phi)\stackrel{\lea}{\d_\beta}+Y_{\beta
}(\phi)\stackrel{\lea}{\d_\al}(-1)^{\ve_\al\ve_\al}](-1)^{\ve_\al},
\e{506}
and
\be
&&{(Y_{\al}(\phi), Y_{\beta}(\phi), Y_{\ga}(\phi))'}^a_{Q(\phi)}
(-1)^{(\ve_\al + 1)(\ve_\ga + 1)}\equiv\nn\\&&
\equiv
 Q^a(\phi) \stackrel{\lea}{\d_\al} \stackrel{\lea}{\d_\beta}
\stackrel{\lea}{\d_\ga} (i
\hbar)^3 (-1)^{\ve_\al
\ve_\ga} =\nn\\&& = {1\over 3} \left([{(Y_{\al}(\phi), Y_{\beta
}(\phi))'}^a_{Q(\phi)}, Y_{\ga}(\phi)]
(-1)^{\ve_\ga + (\ve_\al + 1)(\ve_\ga +
1)} + cycle (\al, \beta,
\ga)\right) +\nn\\&&+ {1\over 3} i \hbar
\left([[Q^a(\phi), Y_{\al}\phi)], Y_{\beta}(\phi)
\stackrel{\lea}{\d_\ga} + Y_{\ga}(\phi) \stackrel{\lea}{\d_\beta}
(-1)^ {\ve_\beta
\ve_\ga}] (-1)^{\ve_\al
\ve_\ga} + cycle (\al, \beta, \ga
)\right) +\nn\\&&+ {1\over 6} (i
\hbar)^2 \left([Q^a(\phi), (Y_{\al}(\phi) \stackrel{\lea}{\d_\beta} +
Y_{\beta } (\phi)\stackrel{\lea}{\d_\al} (-1)^{\ve_\al\ve_\beta})
\stackrel{\lea}{\d_\ga}] (-1)^{\ve_\al
\ve_\ga} + cycle (\al, \beta, \ga
)\right)
\e{507}
 These expressions deviate from \r{201} and \r{205}
by
$\hbar$-terms and $\hbar^2$-terms. Such terms we also had for the ordinary
quantum
antibrackets generated by a nilpotent operator $Q(\phi)$ \cite{OG}. Their
interpretation is the same here. They are the price of a reparametrization
independent extension of their definitions
  onto the space of parameters $\phi^{\al}$. Note that \r{506} and
\r{507} are second and third derivatives of  scalars, which are not
tensors. Thus
only within a preferred coordinate frame and at least at a fixed value of the
parameters
$\phi^{\al}$  can one expect the formulas
\r{506} and \r{507} to reproduce the original
quantum Sp(2)-antibrackets \r{201} and
\r{205}.
 Since we expect the canonical coordinates
 to be the preferred ones,
all antibrackets should be reproduced  at $\phi^{\al} = 0$.
The vanishing of the
$\hbar$-terms  in \r{506} and \r{507} require
the same conditions, while the
vanishing of the
$\hbar^2$-deviation imposes a new condition in \r{507}.
In the corresponding
$\phi$-extended
$n$-antibrackets obtained from \r{504} we expect
to have deviations involving up to
$n-2$ cyclically symmetrized derivatives of the
2-antibracket deviation which in
terms of canonically coordinates should vanish at $\phi^a=0$.
The $\phi$-extended
$n$-antibrackets will then exactly  reproduce the
original $n$-antibrackets at $\phi^a = 0$. Thus, we should have
\be
&&
{(Y_{\al_1}(0),
Y_{\al_2}(0),\ldots,Y_{\al_n}(0))}^a_{Q}=
\nn\\&&={(Y_{\al_1}(\phi),
Y_{\al_2}(\phi),\ldots,
Y_{\al_n}(\phi))'}^a_{Q(\phi)}|_{\phi=0},
\e{508}
where
the left-hand side are the brackets in section 2.
(At $\phi^\al=0$ we have $Y_\al(0)=\theta_\al+
\{\mbox{\small possible ghost
dependent terms}\}$.)
However, note that
\be
&&{(\ldots,T_{\al_k}^{bc}(0),\ldots)}^a_{Q}
\neq{(\ldots,T_{\al_k}^{bc}(\phi),\ldots)'}^a_{Q(\phi)}|_{\phi=0}=0.
\e{5081}
Thus, for brackets involving the operators
$T^{ab}_\al$ the additional $\hbar$-terms
do not vanish at
$\phi^\al=0$. We have
\eg
\be
&&{(T^{ab}_\al, Y_\beta)}^c=\half[T^{ab}_\al,
[Q^c, Y_\beta]]=-\half
i\hbar[Q^c\stackrel{\lea}{\d}_{\beta},
T^{ab}_\al](-1)^{\ve_\al(\ve_\beta+1)}
\e{5082}
from \r{201}, while \r{50311}  yields
\be
&{(T^{ab}_\al, Y_\beta)'}^c&={(T^{ab}_\al,
 Y_\beta)}^c-\half i\hbar[Q^c,
T^{ab}_\al\stackrel{\lea}{\d}_{\beta}]
(-1)^{\ve_\al}=    -\half i\hbar[Q^c,
T^{ab}_\al]\stackrel{\lea}{\d}_{\beta}
(-1)^{\ve_\al}=\nn\\&&=  -\half i\hbar[Q^c,
T^{ab}_\al\stackrel{\lea}{\d_\beta}-(i\hbar)^{-1}[T_\al^{ab},
Y_\beta] \:](-1)^{\ve_{\al}}=     0,
\e{5083}
where the last equality follows from \r{413} (cf.\r{415}).
This makes
the generalized antibrackets \r{5031} considered at
 $\phi^\al=0$ to be analogues of
the Dirac brackets to those in section 2.

The property \r{503} allows us to derive identities like
\be
&&[Q^a(\phi), Q^b(\phi)]\stackrel{\lea}{\d}_{\al}
\stackrel{\lea}{\d}_{\beta}
\stackrel{\lea}{\d}_{\ga}(i\hbar)^{3}(-1)^{\ve_\beta+
(\ve_\al+1)(\ve_\ga+1)}\equiv 0,
\e{509}
which completely control all Jacobi identities.
Eq.\r{509} implies  in particular
\be
&&(Y_\al, (Y_\beta,
Y_\ga)^{\{a}_Q)_Q^{b\}}(-1)^{(\ve_\al+1)(\ve_\ga+1)}+
cycle(Y_\al,Y_\beta,Y_\ga)=
\nn\\&&
=-\half[(Y_\al,Y_\beta,Y_\ga)_Q^{\{a},
Q^{b\}}](-1)^{(\ve_\al+1)(\ve_\ga+1)},
\e{510}
which is identical to \r{204}.

\section{Quantum master equations and
generalized Maurer-Cartan equations in the
Sp(2)-formalism.}
We propose here that the operators
$\Om_{\al a}$ in the integrability conditions of
\r{407} starting with \r{419} are
determined by the master equations
\be
&&(S, S)^a_{\Delta}=i\hbar[\Delta^a, S],
\e{601}
where $\Delta^a$ are extended Sp(2)-charges defined by
\be
&&\Delta^a\equiv\Om^a+j^a\eta^{\al} \pi_\al(-1)^{\ve_\al},\quad
\Delta^{\{a}\Delta^{b\}}=0,\quad [\phi^\al,
\pi_\beta]=i\hbar\del^\al_\beta,
\e{602}
and where $S$ is an extended ghost charge defined by
\be
&&S(\phi,
\eta, j)\equiv G+j^a\eta^{\al}\Om_{\al
a}(\phi)+{1\over4} j^bj^a\eta^{\beta}
\eta^{\al}(-1)^{\ve_\beta}\Om_{\al\beta a
b}(\phi)+
\nn\\&&+{1\over36}j^cj^bj^a\eta^{\ga}\eta^{\beta
}\eta^{\al}(-1)^{\ve_\beta+\ve_\al\ve_\ga}
\Om_{\al \beta  \ga a b c}(\phi)+
\ldots\nn\\&&\ldots+
{1\over (n!)^2}j^{a_n}\cdots j^{a_1}
\eta^{\al_n}\cdots\eta^{\al_1
}(-1)^{(\ve_{\al_2}+\ldots+\ve_{\al_{n-1}}+
\ve_{\al_1}\ve_{\al_n})}\Om_{\al_1\cdots
\al_n a_1\cdots
a_n}(\phi)+\ldots
\e{603}
In \r{602} and \r{603} we have introduced
the even Sp(2)-parameters $j^a$, and new
ghost variables
$\eta^{\al}$, $\ve(\eta^\al)=\ve_\al+1$,
 which also are to  be treated as
parameters. The former parameter $\phi^\al$
is on the other hand turned into an
operator with conjugate momentum $\pi_\al$.
$G$ in \r{603} is the ghost charge operator
in \r{302} and \r{304} in appendix B. Our main
conjecture is that the operators
$\Om_{\al_1\cdots
\al_n a_1\cdots
a_n}(\phi)$ in \r{603} may  be identified with $\Om_{\al a}$,
$\Om_{\al \beta a b}$ in
\r{419} and all the $\Om$'s in their
integrability conditions in a particular manner.
They satisfy the properties ($S$ is an even operator)
\be
&&\ve(\Om_{\al_1\cdots
\al_n a_1\cdots
a_n}(\phi))=\ve_{\al_1}+\ldots+\ve_{\al_n}+
n,\nn\\&&[G, \Om_{\al_1\cdots
\al_n a_1\cdots
a_n}(\phi)]=-n i\hbar\Om_{\al_1\cdots
\al_n a_1\cdots
a_n}(\phi).
\e{604}
The last relation implies that $\Om_{\al_1\cdots
\al_n a_1 \cdots
a_n}(\phi)$ has ghost number minus $n$.
If we assign ghost
number one to $\eta^{\al}$ and ghost number
zero to $j^a$, then $\Delta^a$ has ghost
number one and
$S$ has ghost number zero.   $(S, S)^a_{\Delta}$
 in the master equations \r{601} are
the quantum Sp(2)-antibrackets  defined in
accordance with \r{201}. Thus, we have
\be
&&\quad (S, S)^a_{\Delta}\equiv[[ S, \Delta^a],
S]=[ S, [\Delta^a, S]].
\e{606}
 By consistency the master equations \r{601}
 require  $[\Delta^a, S]$ to satisfy the
same algebra as
$\Om^a$,
\ie
\r{301} in appendix B, since (cf. \r{2021})
\be
&&0=i\hbar[\Delta^{\{a}, [\Delta^{b\}}, S]]=[\Delta^{\{a}, (S,
S)^{b\}}_{\Delta}]= [[\Delta^a, S], [\Delta^b, S]] .
\e{607}
Another property which also follows from
\r{606} is that the master equations \r{601}
may be written as
\be
&&[S, [\Delta^a, S]]=i\hbar[\Delta^a, S],
\e{6071}
which when compared with \r{302} in
appendix B shows that $S$ indeed
is an extended ghost charge and
$[\Delta^a, S]$ extended Sp(2)-charges.
The explicit form of $[\Delta^a, S]$ to the
lowest orders in
$\eta^{\al}$ are
\be
&&[S, \Delta^d]=i\hbar\Om^d+j^a\eta^{\al}[\Om_{\al a},
\Om^d]+j^dj^a\eta^{\beta}\eta^{\al}\Om_{\al
a}\stackrel{\lea}{\d_\beta}i\hbar(-1)^{\ve_\beta}+\nn\\&&+
{1\over 4}j^bj^a\eta^{\beta}\eta^{\al }[\Om_{\al \beta a b},
\Om^d](-1)^{\ve_\beta}+{1\over 4}j^dj^bj^a\eta^{\ga }
\eta^{\beta }\eta^{\al }\Om_{\al
\beta a b}
\stackrel{\lea}{\d_\ga}i\hbar
(-1)^{\ve_\beta+\ve_\ga}+\nn\\&&+{1\over
36}j^cj^bj^a\eta^{\ga }\eta^{\beta }
\eta^{\al }[\Om_{\al \beta  \ga a b c},
\Om^d](-1)^{\ve_\beta+\ve_\al\ve_\ga}+O(\eta^4).
\e{608}
Inserting \r{603} and \r{608}
into the master equations \r{601} we find that they
 are satisfied identically to
zeroth and first order in
$\eta^{\al }$. However, to second order in $\eta^{\al }$ they
yield exactly
\r{419}, and
 to  third order in $\eta^{\al }$ they yield
\be
&&\left(\del_a^d\d_\al\Om_{\beta  \ga
b c}(-1)^{\ve_\al\ve_\ga}+\half(i\hbar)^{-2}(\Om_{\al a},
\Om_{\beta \ga b
c})^d_{\Om}(-1)^{\ve_\al\ve_\ga}+cycle(a, b, c)
\right)+\nn\\&&+cycle(\al, \beta ,
\ga )=\left(-(i\hbar)^{-3}(\Om_{\al \{a},
\Om_{\beta b\}},
\Om_{\ga c})^d_{\Om}(-1)^{\ve_\al\ve_\ga}+cycle(a, b,
c)\right)-\nn\\&&-{2\over3}(i\hbar)^{-1}[\Om'_{\al
\beta  \ga a b c},\Om^d],\nn\\ &&
\Om'_{\al
\beta  \ga a b c}\equiv\Om_{\al
\beta  \ga a b c}-\nn\\&&-{1\over8}
\left\{\left((i\hbar)^{-1}[\Om_{\al  \beta a b},
\Om_{\ga c}](-1)^{\ve_\al\ve_\ga}+
cycle(a, b, c)\right) +cycle(\al, \beta, \ga
)\right\},
\e{609}
where the 3-antibrackets are defined by
\r{205} with $Q^a$ replaced by $\Om^a$.

Comparing equations \r{609} and the
integrability conditions of \r{419} we find exact
agreement. We have also checked that the
 consistency conditions \r{607}  yield
exactly
\r{414} to second order in $\eta^{\al}$,
 which is consistent with \r{419} as they
should. Similarly we have checked that \r{607} to third
order in $\eta^{\al}$  yield  conditions
 which are consistent with \r{609}, exactly
like
\r{414} are consistent with \r{419}.

 The
master equations \r{601} yield at higher
 orders in $\eta^{\al }$ equations
 involving still higher quantum
Sp(2)-antibrackets  and
operators
$\Om_{\al  \beta  \ga a b c\ldots}$ with
 still more indices. We conjecture that these
equations all are consistent  with the integrability conditions of
\r{609}.

Notice that $j^a$ and $\eta^\al$ are parameters in \r{603}.
 $\eta^\al$ play the same
role  as in the Sp(1)-formalism  in \cite{OG},
\ie they select (super)antisymmetric
sector with respect to the Greek subscripts
 in \r{419}, \r{609}, etc  in
order to reproduce the correct chain of
 Maurer-Cartan equations within
the framework of the generating equations
\r{601}. The bosonic parameters
$j^a$ are necessary in order to select just the
symmetric sector with respect to the
Sp(2)-subscripts in all structure
relations \r{419}, \r{609}, etc.,
in accordance with the assertion given in
the phrase before \r{414}.

\section{Some formal properties of quantum
antibrackets and quantum master equations.}
In the classical Sp(2)-formalism there are
first order differential operators, $V^a$,
which play a fundamental role in triplectic
quantization \cite{sp2,Trip1,Trip2}.
The quantum analogues are odd operators
$V^a$ satisfying the properties
\be
&&[Q^{\{a}, V^{b\}}]=0,
\e{701}
\be
&&[V^{\{a}, (f, g)_Q^{b\}}]=([V^{\{a},f], g)^{b\}}_Q+(f,
[V^{\{a},g])^{b\}}_Q(-1)^{\ve_f+1},
\e{702}
\be
&&[V^a, V^b]=0,
\e{703}
where $Q^a$ satisfy \r{202}. When these properties are satisfied
then
$\bar{Q}^a\equiv Q^a+kV^a$ with an arbitrary constant
$k$ also satisfy \r{202} and may
be used instead of $Q^a$ in the definition \r{201} of
quantum antibrackets. (Note that
$Q^a$ satisfy
\r{701}, \r{702}, and \r{703} with $V^a$
 replaced by $Q^a$.) This situation we had
in \r{219} in section 2. A particular solution of \r{701}-\r{703} is
\be
&&V^a\equiv (i\hbar)^{-1}[Q^a, H],
\e{704}
where $H$ is an arbitrary even operator which satisfies the property
\be
&&[Q^{\{a},(H,H)^{b\}}_Q]=0.
\e{705}

Consider the quantum master equations \r{601}, \ie
\be
&&(S, S)^a_{\Delta}=i\hbar[\Delta^a, S],
\quad [\Delta^a, \Delta^b]=0.
\e{706}
If we define $\bar{S}$ and $\bar{\Delta}^a$ by
\be
&&\bar{S}\equiv e^{{i\over\hbar}F}
Se^{-{i\over\hbar}F}, \quad
\bar{\Delta}^a\equiv e^{{i\over\hbar}F}
{\Delta}^a e^{-{i\over\hbar}F},
\e{707}
where $F$ is an arbitrary even operator,
then  $\bar{S}$ satisfy the following master
equations
\be
&&(\bar{S}, \bar{S})^a_{\bar{\Delta}}=
i\hbar[\bar{\Delta}^a, \bar{S}].
\e{708}
If now $F$ in \r{707} also satisfies
the master equations \r{706}, \ie
\be
&&(F, F)^a_{\Delta}=i\hbar[\Delta^a, F],
\e{709}
then $\bar{\Delta}^a$ in \r{707} reduce to
\be
&&\bar{\Delta}^a=\Delta^a-(i\hbar)^{-1}
[\Delta^a, cF]=\Delta^a-cV^a,
\e{710}
where $c\equiv 1-e^{-1}$. Note that \r{709}
implies that $F$ satisfies  condition
\r{705} with $H$ and $Q^a$ replaced by
$F$ and $\Delta^a$ respectively.

There are also transformations on $S$
leaving $\Delta^a$ unaffected for which the
master equations are invariant.
These natural automorphisms exist also for the master
equation given in \cite{OG}. We consider this case first.
The integrability conditions of \r{401} expressed in terms of
$\Om_\al$ in \r{403}  were in \cite{OG}
proposed to be encoded in the master
equation
\be
&&(S, S)_\Delta=i\hbar[\Delta, S],\quad \Delta^2=0,
\e{b1}
where $S$ is the extended ghost charge
\r{603} without Sp(2) indices and
bosonic parameters $j$ \cite{OG}, and
where $\Delta$ is an extended BRST charge given
by
$\Delta=\Om+\eta^\al\pi_\al(-1)^{\ve_\al}$.
The antibracket in \r{b1} is the ordinary
quantum antibracket \r{101}. The natural
automorphism of \r{b1} is
\be
&&S\;\ra\;S'\equiv \exp{\biggl\{-(i\hbar)^{-2}[\Delta,
\Psi]\biggr\}}S\exp{\biggl\{(i\hbar)^{-2}[\Delta,
\Psi]\biggr\}},
\e{b2}
where $\Psi$ is an arbitrary odd operator.
It is easily seen that $S'$ also satisfies
the master equation \r{b1}. For infinitesimal transformations we have
\be
&&\del S=(i\hbar)^{-2}[S, [\Delta, \Psi]],\nn\\
&&\del_{21} S\equiv(\del_2\del_1-
\del_1\del_2)S=(i\hbar)^{-2}[S, [\Delta,
\Psi_{21}]],\nn\\
&&\Psi_{21}=(i\hbar)^{-2}(\Psi_2, \Psi_1)_{\Delta}.
\e{b3}

In the Sp(2) case we have the master equations \r{706}.
In this case there is an automorphism under
\be
&&S\;\ra\;S'\equiv \exp{\biggl\{
-(i\hbar)^{-3}\half\ve_{ab}[\Delta^b,[\Delta^a,
F]]\biggr\}}S\exp{\biggl\{(i\hbar)^{-3}
\half\ve_{ab}[\Delta^b,[\Delta^a,F]]\biggr\}},
\e{b5}
where $F$ is an arbitrary even operator.
 For infinitesimal transformations we have
here
\be
&&\del S=(i\hbar)^{-3}[S,\half\ve_{ab}
[\Delta^b,[\Delta^a,F]]],\nn\\
&&\del_{21}
S\equiv(\del_2\del_1-\del_1\del_2)S=(i\hbar)^{-3}
[S,\half\ve_{ab}[\Delta^b,[\Delta^a,F_{21}]]],\nn\\
&&F_{21}=-(i\hbar)^{-3}\half\ve_{ab}
[[\Delta^b, F_2],[\Delta^a,F_1]].
\e{b6}

The above automorphisms are
very similar in structure to the automorphisms of the
corresponding classical master
 equations in the BV-quantization. Compare \eg in the
Sp(2) case
\r{b5} with the  corresponding classical automorphism  given in
\cite{sp2a}.\\ \\

\noindent
{\bf Acknowledgments}

I.A.B. would like to thank Lars Brink for his very warm hospitality at the
Institute of Theoretical Physics, Chalmers and G\"oteborg University. The
work is partially supported by
INTAS-RFBR grant 95-0829. The work of I.A.B. is also
supported by INTAS grant 96-0308 and by RFBR grants 96-01-00482, 96-02-17314.
I.A.B. and R.M. are thankful to the Royal Swedish Academy of Sciences for
financial support. \\ \\ \\

\def\theequation{\thesection.\arabic{equation}}
\setcounter{section}{1}
\renewcommand{\thesection}{\Alph{section}}
\setcounter{equation}{0}
\noindent
{\Large{\bf{Appendix A}}}\\ \\
{\bf Defining properties of the
conventional classical Sp(2)-antibrackets.}\\ \\
The defining properties of the antibrackets $(f,g)^a$
for functions $f, g$ on a manifold
$\cA$ are \cite{sp2,Trip1,Trip2} (The complete
triplectic quantization requires a
$6n$ dimensional manifold $\cA$.)\\
\\
1) Grassmann parity
\be
&&\ve((f, g)^a)=\ve_f+\ve_g+1.
\e{a9}
2) Symmetry
\be
&&(f, g)^a=-(g, f)^a(-1)^{(\ve_f+1)(\ve_g+1)}.
\e{a10}
3) Linearity
\be
&&(f+g, h)^a=(f, h)^a+(g, h)^a, \quad (\ve_f=\ve_g).
\e{a11}
4) Jacobi identities
\be
&&(f,(g, h)^{\{a})^{b\}}(-1)^{(\ve_f+1)(\ve_h+1)}+cycle(f,g,h)\equiv0.
\e{a12}
5) Leibniz' rule
\be
&&(fg, h)^a=f(g, h)^a+(f, h)^ag(-1)^{\ve_g(\ve_h+1)}.
\e{a13}
6) For any odd/even parameter $\la$ we have
\be
&&(f, \la)^a=0\quad {\rm any}\;f\in\cA.
\e{a14}
The
most general form of the Sp(2)-antibrackets are explicitly given by \r{216}
\cite{Trip1,Trip2}.\\ \\ \\

\setcounter{section}{2}
\setcounter{equation}{0}
\noindent
{\Large{\bf{Appendix B}}}\\ \\
{\bf Constraints in arbitrary involutions embedded in Sp(2)-charges.}\\ \\
Constraints in arbitrary involutions may at least for finite number of
degrees of
freedom always be embedded in one single, odd, hermitian and nilpotent BFV-BRST
charge $\Om$ provided one introduces ghost operators to the
constraints \cite{BFV}. It is also possible to embed the constraints in
{\em two}
odd, hermitian charges $\Om^a$, $(a=1,2)$, satisfying \cite{sp2ham} (cf \r{202})
\be
&&\Om^{\{a}\Om^{b\}}\equiv [\Om^{a}, \Om^{b}]=0.
\e{301}
This may be done in such a way that there also exists an even, hermitian ghost
charge $G$ satisfying
\be
&&[G, \Om^a]=i\hbar\Om^a.
\e{302}
Explicitly this may be done  by
means of the Sp(2)-covariant ghost operators
$\ca^{\al a}$ in the case $\theta_\al$ are constraint operators
\cite{sp2ham}. Let
$\ca^{\al a}$ and their conjugate momentum operators
$\pet_{\al a}$ satisfy the properties
 \be
&&[\ca^{\al a},
\pet_{\beta b}]=i\hbar\del^\al_\beta\del^a_b,\quad(\cC^{\al
a})\dagg=\cC^{\al a},
\quad\pet_{\al a}\dagg=-(-1)^{\varepsilon_\al}\pet_{\al a},\nn\\
&&\ve(\cC^{\al a})=\ve(\pet_{\al a})=\ve_\al+1, \quad
\ve_\al\equiv\ve(\theta_\al).
\e{303}
The so called new ghost operator in \r{302} is then defined by
\be
&&G\equiv-\half\biggl(\pet_{\al a}\cC^{\al a}-\cC^{\al a}\pet_{\al
a}(-1)^{\ve_\al}\biggr).
\e{304}
The Sp(2) charges $\Om^a$ satisfying \r{301} and \r{302} are then of the form
\be
&&\Om^a=\theta_\al\ca^{\al a}+\half\ca^{\beta b}\ca^{\al a} {U}_{\al
\beta}^{\;\;\;\ga}\pet_{\ga b}(-1)^{\ve_\beta+\ve_\ga}+\cdots ,
\e{305}
where the dots denote terms which are of higher orders in the ghost momenta
$\pet_{\al a}$. ${U}_{\al
\beta}^{\;\;\;\ga}$ are the structure operators in the involution
relations of $\theta_\al$, \ie
\be
&&[\theta_\al, \theta_\beta]=i\hbar{U}_{\al
\beta}^{\;\;\;\ga}\theta_\ga.
\e{306}
(In general these constraint operators $\theta_\al$
are not hermitian.)

The complete Sp(2) formalism also involves Lagrange multipliers $\la^\al$
($\ve(\la^\al)=\ve_\al$) and their conjugate momenta $\pi_\al$ ($[\la^\al,
\pi_\beta]=i\hbar\del^\al_\beta$) \cite{sp2ham}. However, in distinction to
standard
BRST formalism, these Lagrange multipliers are considered as ghost
variables in the
Sp(2) case. In fact, $\la^\al$ has new ghost number two which means that $G$ in
\r{304} then should be replaced by
\be
&&G\equiv-\half\biggl(\pet_{\al a}\cC^{\al a}-\cC^{\al a}\pet_{\al
a}(-1)^{\ve_\al}\biggr)-\biggl(\pi_\al\la^\al+
\la^\al\pi_\al(-1)^{\ve_\al}\biggr).
\e{307}
The Sp(2) charges $\Om^a$ including the Lagrange multipliers are then explicitly
\be
&&\Om^a=\theta_\al\ca^{\al a}+\half\ca^{\beta b}\ca^{\al a} {U}_{\al
\beta}^{\;\;\;\ga}\pet_{\ga
b}(-1)^{\ve_\beta+\ve_\ga}+\nn\\&&+\ve^{ab}\pet_{\beta b}\la^\beta+\half
\la^\beta\ca^{\al a} U^{\;\;\;\ga}_{\al\beta}\pi_\ga+\cdots ,
\e{308}
where the dots denote terms of  order square and higher in the ghost momenta
$\pet_{\al a}$ and/or $\pi_\al$. The  results in sections 3-6 are valid both for
\r{305} and
\r{308}. Note also that $\Om^a$ may be used to construct quantum
Sp(2)-antibrackets.
\\ \\ \\

\end{document}